\documentclass[11pt,a4paper]{article}
\pdfoutput=1
\usepackage{jheppub}
\usepackage{amsmath,amssymb,bm}
\usepackage{graphicx}
\usepackage{hepunits}
\usepackage{booktabs}

\usepackage[utf8]{inputenc}

\allowdisplaybreaks

\usepackage{color}

\title{NNLL momentum-space threshold resummation in direct top quark production at the LHC}

\author[a,b,c]{Li Lin Yang,}
\author[a,c]{Chong Sheng Li,}
\author[d]{Jun Gao}
\author[e]{and Jian Wang}

\affiliation[a]{School of Physics and State Key Laboratory of Nuclear Physics and Technology, Peking University,\\
  Beijing 100871, China}
\affiliation[b]{Collaborative Innovation Center of Quantum Matter, Beijing, China}
\affiliation[c]{Center for High Energy Physics, Peking University, Beijing 100871, China}
\affiliation[d]{Department of Physics, Southern Methodist University,\\
  Dallas, TX 75275-0175, USA}
\affiliation[e]{PRISMA Cluster of Excellence \& Mainz Institute for Theoretical Physics,\\
  Johannes Gutenberg University, D-55099 Mainz, Germany}

\emailAdd{yanglilin@pku.edu.cn}
\emailAdd{csli@pku.edu.cn}
\emailAdd{jung@mail.smu.edu}
\emailAdd{jian.wang@uni-mainz.de}

\abstract{We update the theoretical precision of the total cross section for direct top quark production at the LHC by extending the threshold resummation to the next-to-next-to-leading logarithmic accuracy.}


\arxivnumber{1409.6959}
\preprint{MITP/14-068}

\begin{document}

\maketitle

\section{Introduction}
\label{sec:introduction}

The CERN Large Hadron Collider (LHC) has finished operating at center-of-mass energies of $\unit{7}{\TeV}$ and $\unit{8}{\TeV}$. After a major upgrade phase, the machine is expected to run again in 2015. The total delivered integrated luminosity has reached $\unit{27}{\invfb}$ for both the ATLAS and CMS detectors. Successful run of the LHC has resulted in the discovery of a Higgs boson, whose properties are consistent with the standard model predictions \cite{Aad:2012tfa, Chatrchyan:2012ufa}.

The high luminosity as well as the excellent performance of the detectors ensure us to measure various standard model processes with highest precisions ever attained. In particular, the knowledges about the processes involving top quarks have been greatly improved by the new LHC data, and will certainly be further improved in the near future. With only an integrated luminosity of up to \unit{4.9}{\invfb} collected from the run at \unit{7}{\TeV}, the LHC has already provided a combined measurement of the top quark mass $m_t = \unit{$173.29 \pm 0.95$}{\GeV}$ \cite{ATLAS-CONF-2013-102, CMS-PAS-TOP-13-005}, whose uncertainty is comparable to the latest Tevatron combination $m_t = \unit{$174.34 \pm 0.64$}{\GeV}$ \cite{Tevatron:2014cka}. A first combination of the Tevatron and LHC measurements gave rise to $m_t = \unit{$173.34 \pm 0.76$}{\GeV}$ \cite{ATLAS:2014wva}. Besides the top quark mass, the total production cross sections as well as various differential distributions have also been measured with high precisions (see, e.g., report by Tancredi Carli at ICHEP 2014).

On the theory side, predictions for the production and decay processes of top quarks are also greatly improved in the recent years with higher order QCD corrections available. The next-to-next-to-leading order (NNLO) QCD corrections to the total cross sections for the top quark pair production have been obtained in \cite{Baernreuther:2012ws, Czakon:2012zr, Czakon:2012pz, Czakon:2013goa}. Predictions for the differential distributions in top quark pair production are available with soft gluon resummation at next-to-next-to-leading logarithmic (NNLL) accuracy, combined with next-to-leading order (NLO) fixed order results \cite{Ahrens:2010zv, Ahrens:2011mw,Zhu:2012ts,Broggio:2014yca}. The single top quark production processes are also known with similar precisions \cite{Kidonakis:2010tc,Zhu:2010mr,Kidonakis:2011wy,Wang:2012dc}. Finally, the calculation of the fully differential top quark decay rates at NNLO in QCD was performed in \cite{Gao:2012ja, Brucherseifer:2013iv}.

Based on the reliable theoretical predictions and precise experimental measurements, the top quark can serve as an excellent probe for new physics beyond the standard model. In this paper, we will deal with a particular type of new physics, which contributes to the flavor-changing neutral interactions of the top quark. These contributions can be parameterized as a model-independent effective theory into several anomalous flavor-changing couplings $\kappa_{tqZ}$, $\kappa_{tq\gamma}$, $\kappa_{tqH}$ and $\kappa_{tqg}$, where $q$ stands for up or charm quark and $g$ denotes gluon. These anomalous couplings can be probed in the productions and decays of top quarks, and have already been studied by the ATLAS and CMS collaborations at the LHC \cite{Aad:2012gd, Aad:2012ij, Chatrchyan:2012hqa, Chatrchyan:2013nwa, CMS-PAS-TOP-12-037, ATLAS-CONF-2013-063, CMS-PAS-HIG-13-034, CMS-PAS-TOP-14-003, CMS-PAS-TOP-14-007}.
We will concentrate on the anomalous couplings involving the gluon, whose most stringent exclusion limits come from the measurement in \cite{ATLAS-CONF-2013-063} and are given by
\begin{align}
  \label{eq:bounds_kappa}
  \kappa_{tug}/\Lambda < \unit{0.0051}{\TeV^{-1}} \, , \quad \kappa_{tcg}/\Lambda < \unit{0.011}{\TeV^{-1}} \, ,
\end{align}
at the 95\% confidence level (C.L.). Note that the above limits apply to the anomalous couplings at the electroweak scale (more precisely, at the scale $m_t$), under the assumption that only one non-zero coupling is present at a time. On the other hand, if one computes the effective couplings in a specific new physics model, the intrinsic scale for them would be the cut-off scale $\Lambda$. The couplings at the two scales, $\kappa_{tqg}(m_t)$ and $\kappa_{tqg}(\Lambda)$, are related by renormalization group (RG) equations, involving possible mixings from other operators. In this paper, we work in the low energy effective theory, and take the couplings at $m_t$ as input. Therefore, we do not need to consider the RG evolution of the operators down from the new physics scale $\Lambda$.

The constraints on the anomalous couplings in Eq.~(\ref{eq:bounds_kappa}) can be translated to constraints on the branching ratios of the top quark rare decays using the NLO QCD results in \cite{Zhang:2008yn}, which reads
\begin{align}
  \label{eq:bounds_br}
  \text{BR}(t \to u g) < 3.1 \times 10^{-5} \, , \quad \text{BR}(t \to c g) < 1.4 \times 10^{-4} \, .
\end{align}
It is instructive to compare these constraints with the standard model (SM) predictions. In the standard model, the above decays are forbidden at tree-level and can only arise starting from one-loop diagrams with a $W$ boson exchanged. Such contributions are highly suppressed due to the GIM mechanism \cite{Glashow:1970gm}. The resulting branching ratios are given by \cite{AguilarSaavedra:2004wm}
\begin{align}
  \text{BR}_{\text{SM}}(t \to u g) \approx 4 \times 10^{-14} \, , \quad \text{BR}_{\text{SM}}(t \to c g) \approx 5 \times 10^{-12} \, ,
\end{align}
which are orders of magnitudes smaller than the current experimental upper bounds and are beyond the reach of the LHC. As a consequence, if the LHC measures any top quark flavor-changing neutral process in the near future, it will be an definitive indication of new physics beyond the standard model.

In extracting the limits in Eq.~(\ref{eq:bounds_kappa}), the theoretical predictions for the direct top quark production process in \cite{Liu:2005dp, Gao:2011fx} based on NLO QCD were used. Several related processes were also calculated to the same precision. For example, The NLO QCD corrections to the decays of the top quark through flavor-changing neutral interactions have been calculated in \cite{Zhang:2008yn, Zhang:2010bm, Drobnak:2010wh, Drobnak:2010by}; the NLO QCD corrections to the production of a single top quark associated with an up or charm quark were also calculated in \cite{Gao:2011fx}. Among these processes, the direct top quark production is the most sensitive to the anomalous couplings. For this process, an improved prediction combining soft gluon resummation at next-to-leading logarithmic (NLL) accuracy with NLO fixed order results is available in \cite{Yang:2006gs,Kidonakis:2014dua}.

There are several reasons to go beyond this accuracy for direct top quark production. The first is that with the LHC running at a higher energy and delivering much more integrated luminosity, the experimental measurements of the anomalous couplings can be largely improved. This calls for a better theoretical understanding of the relevant observables. The second reason is that the current techniques allow people to perform the threshold resummation at next-to-next-to-leading logarithmic (NNLL) accuracy for a broad class of processes. Full next-to-next-to-leading order (NNLO) calculations are also available for several complicated processes beyond the simple ones like Drell-Yan or Higgs boson production. In view of these developments, it is then attempting to apply these techniques also to processes beyond the standard model. Finally, we have found that an inconsistent treatment of the power corrections was used in \cite{Yang:2006gs}. While it is not wrong since the resummation formula is formally valid only at leading power, this has led to visible changes in the numerical results. Therefore we would like to update those results with a more consistent prescription for the power corrections.

This paper is organized as following. In Sec.~\ref{sec:framework} we present our resummation framework which serves as the basis for later phenomenological studies. In Sec.~\ref{sec:results} we show and discuss the updated numerical predictions for the total cross section. We conclude in Sec.~\ref{sec:conclusion}.

\section{Framework}
\label{sec:framework}

We will consider effective operators of the form
\begin{align}
  \mathcal{L}_{\text{eff}} = -g_s \sum_{q=u,c} \frac{\kappa_{tqg}}{\Lambda} \, \bar{t}\sigma^{\mu\nu}T^aq \, G^a_{\mu\nu} + \mathrm{h.c.},
\end{align}
where $\kappa_{tqg}$ are the anomalous couplings, $\Lambda$ is the new physics scale, $T^a$ are the $SU(3)$ generators, and $G^a_{\mu\nu}$ are the gluon field strength tensors. We have suppressed the possible chiral structure in the above operators, which has no effect on the total cross sections discussed in this paper.

At leading order in QCD, these operators will induce direct top quark production processes
\begin{align}
  g(p_1) + q(p_2) \to t(k) \, ,
\end{align}
where $q=u,c$. It is obvious that an anti-top quark may also be produced with $q=\bar{u},\bar{c}$. The total cross section for direct top quark production at the LHC can be written as a convolution of perturbative partonic cross sections with parton distribution functions (PDFs):
\begin{align}
	\label{eq:hadcs}
  \sigma(\tau,m_t) = \sum_{i,j} \tau \int dx_1 dx_2 dz \, \delta(\tau-x_1x_2z) \, f_{i/p}(x_1,\mu_f) \, f_{j/p}(x_2,\mu_f) \, \frac{\hat{\sigma}_{ij}(z,m_t,\mu_f)}{z} \, ,
\end{align}
where $\tau=m_t^2/s$ with $\sqrt{s}$ being the center-of-mass energy of the collider and $m_t$ being the top quark mass. The partonic center-of-mass energy is given by $\sqrt{\hat{s}}$ with $\hat{s}=(p_1+p_2)^2=x_1x_2s$. The variable $z$ is then $m_t^2/\hat{s}$. The partonic cross sections can be written as a series in the strong coupling constant $\alpha_s$:
\begin{align}
  \hat{\sigma}_{ij}(z,m_t,\mu_f) &\equiv z \, C_{ij}(z,m_t,\mu_f) = \frac{8\pi^2\alpha_s(\mu_r)}{3} \left( \frac{\kappa(\mu_r)}{\Lambda} \right)^2 z \sum_n \left( \frac{\alpha_s}{4\pi} \right)^n C_{ij}^{(n)}(z,m_t,\mu_r,\mu_f) \, .
\end{align}
Note that the renormalization scale dependence of $\kappa$ through its RG equation (see Appendix~\ref{app:a}) is required to ensure the scale-independence of the cross sections. Since we consider only one non-zero anomalous coupling at a time, operator mixing has no impact here.

At leading order, it is clear that $C_{gq}^{(0)}=C_{qg}^{(0)}=\delta(1-z)$, and $C_{ij}^{(0)}=0$ otherwise. The next-to-leading order results of the partonic cross sections have been calculated in \cite{Liu:2005dp}, which were recalculated in \cite{Gao:2011fx} to study the sensitivity of the LHC for this process. It was then found out that there are several typos in the expressions in \cite{Liu:2005dp}. In spite of that, the numerical results were found to be almost unaffected. The correct expressions were used in \cite{Gao:2011fx}. However, they are too long to be put in a letter-type article, and therefore we list them below for the convenience of the readers.
\begin{align}\label{eqs:c}
  C_{gq}^{(1)}(z,m_t,\mu_r,\mu_f) &= \frac{2}{3} \, \Biggl\{ \left[ - 8\ln\frac{\mu_f^2}{m_t^2} - \frac{19}{2}\ln\frac{\mu_f^2}{\mu_r^2} - 15 + \frac{17\pi^2}{3} \right] \delta(1-z) \nonumber
  \\
  &\quad + 26 \left[ \frac{1}{1-z} \ln \left(\frac{m_t^2(1-z)^2}{\mu_f^2z} \right) \right]_+ - \frac{8}{(1-z)_+} \nonumber
  \\
  &\quad + \left( \frac{18}{z} - 40 + 14z - 18z^2 \right) \ln \left(\frac{m_t^2(1-z)^2}{\mu_f^2z} \right) \nonumber
  \\
  &\quad + \left( -\frac{2}{1-z} + 6 - 8z + 10z^2 \right) \ln z - \frac{21}{2z} + 15 - \frac{25z}{2} + 14z^2 \Biggr\} \, , \nonumber
  \\
  C_{gg}^{(1)}(z,m_t,\mu_r,\mu_f) &= 4 \, \Biggl\{\ln \left( \frac{m_t^2(1-z)^2}{\mu_f^2z} \right) \frac{z^2+(1-z)^2}{2} \nonumber
  \\
  &\hspace{5em} + \left[ \frac{3}{4} + z + \frac{5z^2}{4} - \frac{1}{4(1+z)} \right] \ln z + \frac{3}{4z} - \frac{3}{8} + \frac{59z}{16} - \frac{65z^2}{16} \Biggr\} \, , \nonumber
  \\
  C_{qq}^{(1)}(z,m_t,\mu_r,\mu_f) &= \frac{16}{3} \, \Biggl\{ \ln \left(\frac{m_t^2(1-z)^2}{\mu_f^2z} \right) \frac{1+(1-z)^2}{z} - \frac{5}{3z} + \frac{8}{3} \Biggr\} \, , \nonumber
  \\
  C_{q\bar{q}}^{(1)}(z,m_t,\mu_r,\mu_f) &= \frac{8}{3} \, \Biggl\{ \ln \left(\frac{m_t^2(1-z)^2}{\mu_f^2z} \right) \frac{1+(1-z)^2}{z} - \frac{2}{z} + \frac{11}{3} - \frac{4z}{3} + \frac{2z^2}{3} \Biggr\} \, , \nonumber
  \\
  C_{qq'}^{(1)}(z,m_t,\mu_r,\mu_f) &= \frac{8}{3} \, \Biggl\{ \ln \left(\frac{m_t^2(1-z)^2}{\mu_f^2z} \right) \frac{1+(1-z)^2}{z} - \frac{2}{z} + 3 \Biggr\} \, , \nonumber
  \\
  C_{q'\bar{q}'}^{(1)}(z,m_t,\mu_r,\mu_f) &= \frac{8}{3} \, \Biggl[ \frac{1}{3z} - z + \frac{2z^2}{3} \Biggr] \, .
\end{align}
In the above formulas, $q$ stands for the quark flavor entering the leading order process, $\bar{q}$ represents the antiparticle of $q$, and $q'$ denotes all other flavors of quarks and antiquarks. The typos in \cite{Liu:2005dp} happened to be in $C_{qq}^{(1)}$ and $C_{q\bar{q}}^{(1)}$, and we have checked that the differences lead to nearly no numerical impacts.

All partonic cross sections are regular functions of $z$ except $C_{gq}$, which contains $\delta(1-z)$ as well as singular plus-distributions of the form
\begin{align}
  P_n(z) \equiv \left[ \frac{1}{1-z} \ln^n \left( \frac{m_t^2(1-z)^2}{\mu^2z} \right) \right]_+ ,
\end{align}
where $n=0$ or $1$ at the NLO. At higher orders in perturbation theory, accompanied with each power of $\alpha_s$, the maximal value of the power $n$ will be increased by $2$. If the hadronic cross section is saturated by the region $z \lesssim 1$, these terms will lead to large corrections resulting in poor perturbative convergence. A resummation of these distributions to all orders in $\alpha_s$ can be achieved utilizing the Mellin or Laplace transforms, which are defined by
\begin{align}
  \mathcal{M}[f(z)] = \int_0^1 dz \, z^{N-1} \, f(z) \, , \quad \mathcal{L}[f(z)] = \int_0^\infty d\xi \, e^{-\xi N} f(z) \, ,
\end{align}
respectively, where $\xi=(1-z)/\sqrt{z}$, $N$ is called the moment conjugate to $z$. The limit $z \to 1$ corresponds to the limit $N \to \infty$ in moment space, and it can be shown that Mellin transform and Laplace transform are actually equivalent in this limit, i.e., $\mathcal{M}[f(z)] - \mathcal{L}[f(z)] = \mathcal{O}(1/N)$. The transform rules for the delta function and the plus-distributions are listed in Table~\ref{tab:transform}, where $\bar{N}=Ne^{\gamma_E}$ with $\gamma_E$ the Euler constant.

\begin{table}[ht!]
  \centering
  \begin{tabular}{ccc}
    \toprule
    $f(z)$ & $\mathcal{M}[f(z)]$ & $\mathcal{L}[f(z)]$
    \\
    \midrule
    $\delta(1-z)$ & 1 & 1
    \\
    $P_0(z)$ & $-\ln\bar{N} + \mathcal{O}(1/N)$ & $-\ln\bar{N}$
    \\
    $P_1(z)$ & $\displaystyle \ln^2\bar{N} - \ln\bar{N}\ln\frac{m_t^2}{\mu^2} + \mathcal{O}(1/N)$ & $\displaystyle \ln^2\bar{N} - \ln\bar{N}\ln\frac{m_t^2}{\mu^2}$
    \\
    \bottomrule
  \end{tabular}
  \caption{Mellin and Laplace transforms of the delta function and the plus-distributions.}
  \label{tab:transform}
\end{table}

In the moment space and in the limit $N \to \infty$, all partonic cross sections become power suppressed by $1/N$ except $C_{gq}=C_{qg}$, which becomes
\begin{align}
	\label{eq:leading}
  \tilde{c}_{gq}(N,m_t,\mu_f) &= \tilde{c}^{\text{leading}}_{gq}(N,m_t,\mu_f) + \mathcal{O}(1/N) \nonumber
  \\
  &= \frac{8\pi^2\alpha_s}{3} \left( \frac{\kappa}{\Lambda} \right)^2 \left\{ 1 + \frac{\alpha_s}{\pi} \left[ \frac{13}{3} \ln^2\bar{N} + \ln\bar{N} \left( \frac{13}{3} \ln\frac{\mu_f^2}{m_t^2} + \frac{4}{3} \right) \right. \right. \nonumber
  \\
  &\hspace{4em} \left. \left. - \frac{4}{3} \ln\frac{\mu_f^2}{m_t^2} - \frac{19}{12} \ln\frac{\mu_f^2}{\mu_r^2} - \frac{5}{2} + \frac{17\pi^2}{18} \right] + \mathcal{O}(\alpha_s^2) \right\} + \mathcal{O}(1/N) \, .
\end{align}
It has been shown in \cite{Yang:2006gs} that the above partonic cross section can be factorized into the product of a hard function and a soft function
\begin{align}
  \tilde{c}_{gq}(N,m_t,\mu_f) = H(m_t,\mu_f) \, \tilde{s}(L,\mu_f) + \mathcal{O}(1/N) \, ,
\end{align}
where $L = \ln(m_t^2/\mu_f^2\bar{N}^2)$. Up to the next-to-leading order, the hard and soft functions can be extracted from the expressions in \cite{Yang:2006gs} and are given by
\begin{align}\label{eqs:hard_function}
  H(m_t,\mu) &= \frac{8\pi^2\alpha_s}{3} \left( \frac{\kappa}{\Lambda} \right)^2 \left[ 1 + \frac{\alpha_s}{4\pi} \left( -\frac{13}{3} \ln^2\frac{\mu^2}{m_t^2} - 8\ln\frac{\mu^2}{m_t^2} - \frac{46}{3} + \frac{55\pi^2}{18} \right) + \mathcal{O}(\alpha_s^2) \right] ,
  \\
  \tilde{s}(L,\mu) &= 1 + \frac{\alpha_s}{4\pi} \left( \frac{13}{3} L^2 - \frac{8}{3} L + \frac{16}{3} + \frac{13\pi^2}{18} \right) + \mathcal{O}(\alpha_s^2) \, ,
\end{align}

The hard and soft functions satisfy RG equations
\begin{align}
  \frac{d}{d\ln\mu} H(m_t,\mu) &= \left( \frac{13}{3} \gamma_{\text{cusp}}(\alpha_s) \ln\frac{m_t^2}{\mu^2} + 2\gamma^H \right) H(m_t,\mu) \, ,
  \\
  \frac{d}{d\ln\mu} \tilde{s} \bigg( \ln\frac{Q^2}{\mu^2}, \mu \bigg) &= \left( -\frac{13}{3} \gamma_{\text{cusp}}(\alpha_s) \ln\frac{Q^2}{\mu^2} - 2\gamma^S \right) \tilde{s} \bigg( \ln\frac{Q^2}{\mu^2}, \mu \bigg) \, ,
\end{align}
where $Q$ is an arbitrary scale. The resummation of threshold logarithms then follows simply by choosing appropriate scales separately for the hard and soft functions, and then using the above RG equations to evolve the two functions to the same scale. In \cite{Yang:2006gs}, the one-loop anomalous dimensions were used in the RG equations, corresponding to an NLL accuracy. The recent development allows us to use higher-order anomalous dimensions, thus to achieve NNLL accuracy for the resummation. These anomalous dimensions are listed in Appendix~\ref{app:a}. The last step for the resummation amounts to an inverse Mellin or Laplace transform back to the momentum space
\begin{align}
  C_{gq}^{\text{NNLL}}(z) = \mathcal{M}^{-1}\Big[\tilde{c}_{gq}^{\text{NNLL}}(N)\Big]  \quad \text{or} \quad C_{gq}^{\text{NNLL}}(z) = \mathcal{L}^{-1}\Big[\tilde{c}_{gq}^{\text{NNLL}}(N)\Big] \, .
\end{align}

We now discuss the choice of the hard and soft scales. For the hard function, there is not much ambiguity in choosing the hard scale, which should naturally be around $m_t$. The choice of the soft scale, on the other hand, is a bit more subtle. One possibility is to choose it to be around $m_t/N$ in the moment space, as was done in \cite{Yang:2006gs} and many earlier works on threshold resummation. In this way, the inverse Mellin or Laplace transform has to be performed numerically, and one needs some prescription to deal with the Landau pole problem such as the Minimal Prescription \cite{Catani:1996yz} or the Borel Prescription \cite{Abbate:2007qv}. Another possibility, advocated in \cite{Becher:2006nr, Becher:2007ty}, is to choose the soft scale directly in the momentum space, which then corresponds to an ``average'' energy of soft emissions. The consequences of the different choices have been carefully elaborated and compared in the literature \cite{Becher:2007ty, Ahrens:2008nc, Bonvini:2012az, Sterman:2013nya, Almeida:2014uva, Bonvini:2014qga}. One of the benefits of the momentum space formalism is that the inverse Laplace transform can be carried out analytically, bypassing the Landau pole problem. Below, we will perform the NNLL resummation using the momentum space formalism. The NNLL resummation in the moment space formalism as well as a full NNLO calculation are beyond the scope of this paper, and are left to a future study.

Solving the RG equations for the hard and soft functions and transforming back to the momentum space, we can write the resummed partonic cross section as
\begin{align}
 C_{gq}^{\text{NNLL}}(z,m_t,\mu_f) &= H(m_t,\mu_h) \, U(m_t,\mu_h,\mu_s,\mu_f) \, \tilde{s} \left( \ln\frac{m_t^2}{\mu_s^2} + \partial_\eta , \mu_s \right) \frac{z^{-\eta}}{(1-z)^{1-2\eta}} \, \frac{e^{-2\gamma_E\eta}}{\Gamma(2\eta)} \, ,
\end{align}
where $\eta=2a_\Gamma(\mu_s,\mu_f)$ and
\begin{align}
  U(m_t,\mu_h,\mu_s,\mu_f) &= \exp \left[ 4S(\mu_h,\mu_s) - 2a_\Gamma(\mu_h,\mu_s)
    \ln\frac{m_t^2}{\mu_h^2} - 2a_{\gamma^H}(\mu_h,\mu_f) + 2a_{\gamma^S}(\mu_s,\mu_f)
  \right] .
\end{align}
The functions $S(\mu,\nu)$ and $a_{\gamma}(\mu,\nu)$ are defined as \cite{Becher:2006mr}
\begin{align}
  S(\mu,\nu) &= -  \int_{\alpha_s(\mu)}^{\alpha_s(\nu)} d\alpha \, \frac{\Gamma (\alpha)}{\beta(\alpha)} \int_{\alpha_s(\mu)}^{\alpha} \frac{d\alpha^{\prime}}{\beta(\alpha^{\prime})} \, ,
  \\
  a_{\gamma}(\mu,\nu) &= -\int_{\alpha_s(\mu)}^{\alpha_s(\nu)} d\alpha \, \frac{\gamma(\alpha)}{\beta(\alpha)} \, ,
\label{eqs:anoa}
\end{align}
with $\Gamma(\alpha_s) = 13\gamma_{\text{cusp}}(\alpha_s)/6$.
Finally, we can add back the power-suppressed contributions at the NLO by a matching procedure, and hence obtain our prediction at the NLO+NNLL accuracy:
\begin{align}
  \sigma^{\text{NLO+NNLL}} = \sigma^{\text{NNLL}} + \sigma^{\text{NLO}} - \sigma^{\text{NLO,leading}} \, .
\end{align}
This formula will be the starting point of our numerical results presented in the next section.

\section{Numerical results}
\label{sec:results}

\begin{figure}[t!]
\begin{center}
\includegraphics[width=0.45\textwidth]{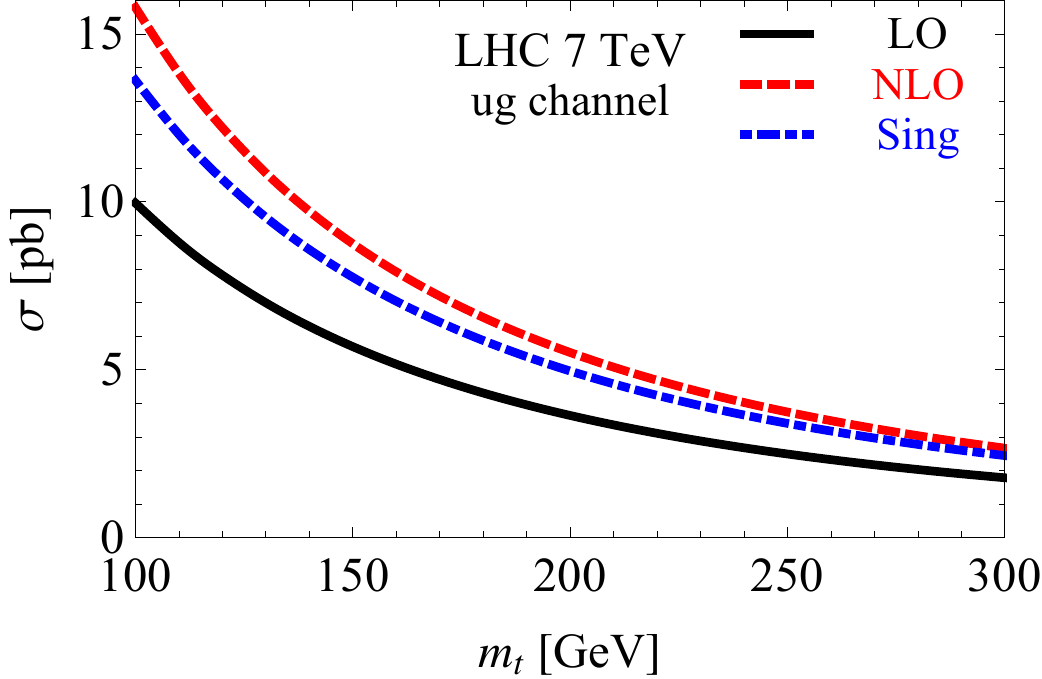}
\includegraphics[width=0.45\textwidth]{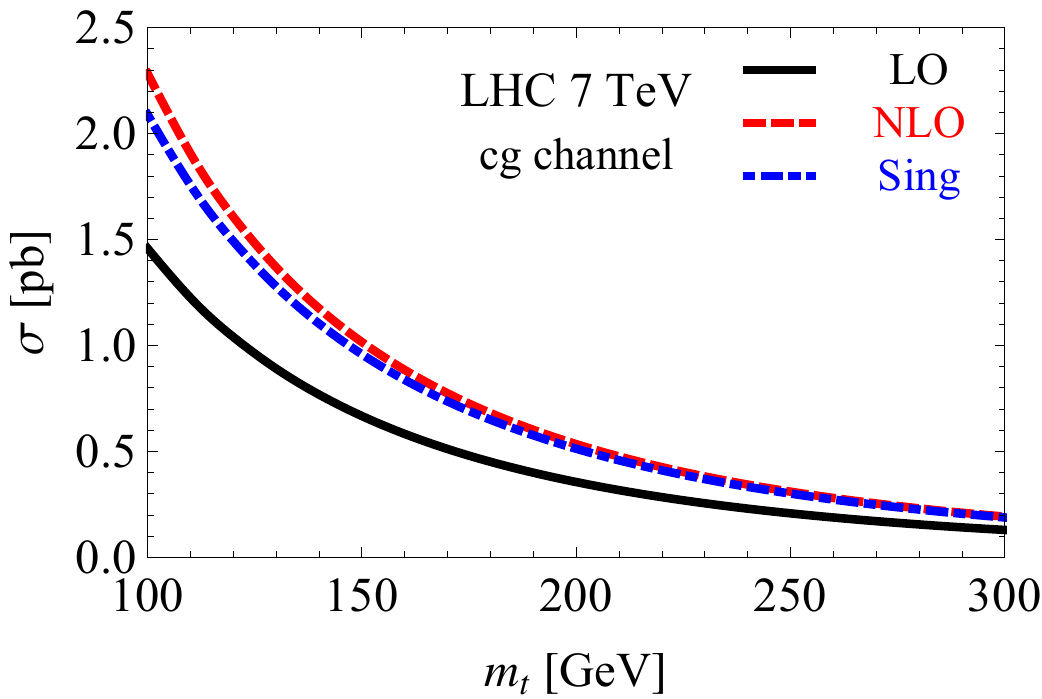}
\\ \vspace{1ex}
\includegraphics[width=0.45\textwidth]{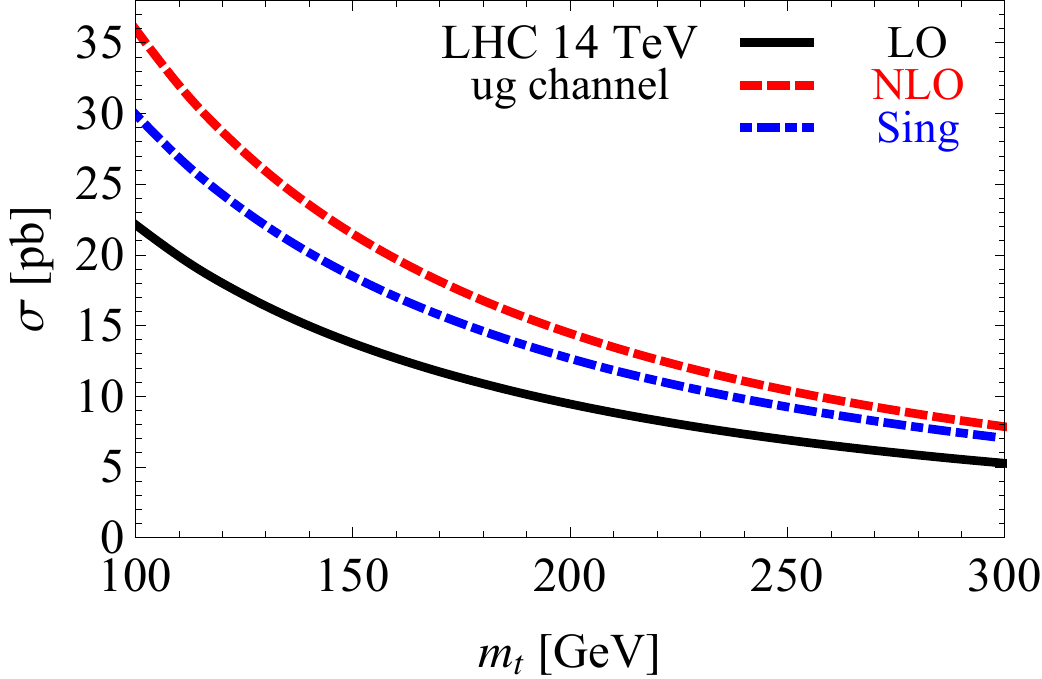}
\includegraphics[width=0.45\textwidth]{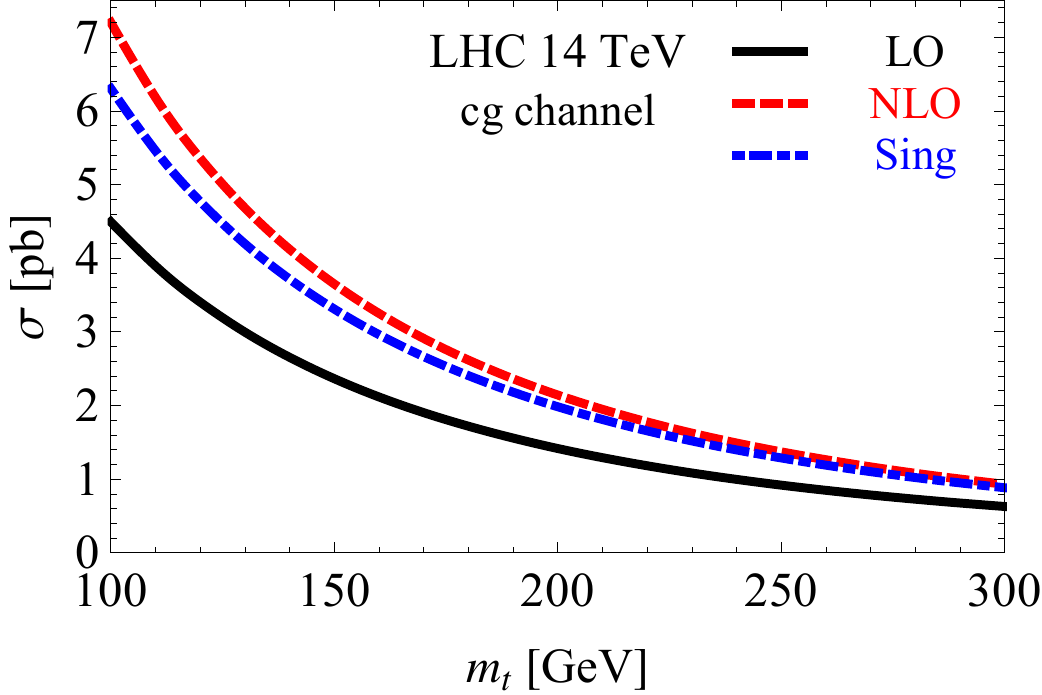}
\end{center}
\vspace{-4ex}
\caption{\label{fig:singular}The LO, NLO and leading singular terms in the $ug$ channel (left) and the $cg$ channel (right) at the LHC.}
\end{figure}

In this section, we present the numerical predictions from our NLO+NNLL resummed results for direct top quark production at the LHC. We choose the top quark mass to be $m_t=\unit{173.2}{\GeV}$. For the anomalous couplings, we adopt the convention that only one of the two couplings is non-zero at a time, and the value is taken to be $\kappa_{tqg}/\Lambda=\unit{0.01}{\TinveV}$. Results for other values of the anomalous couplings can be easily obtained by a simple rescaling. Throughout the numerical calculations, we use the CT10 PDF sets \cite{Lai:2010vv, Gao:2013xoa} and the associated values for the strong coupling constant $\alpha_s$ unless otherwise specified. The default factorization scale $\mu_f$ and renormalization scale $\mu_r$ are chosen as the top quark mass $m_t$.

Before presenting the resummed predictions, we first examine the importance of the leading singular terms defined in Eq.~(\ref{eq:leading}). In Fig.~\ref{fig:singular}, we show the contribution from the leading singular terms as well as the LO and NLO results as a function of the top quark mass. Since the leading singular terms come from soft gluon emissions, one can expect that they become more important for larger top quark masses or smaller center-of-mass energies. We show this feature in Fig.~\ref{fig:singular}, from which one can see that the contribution from the leading singular terms approach the full NLO results for both $\sqrt{s}=\unit{7}{\TeV}$ and $\sqrt{s}=\unit{14}{\TeV}$ when the top quark becomes heavier and heavier. One can also see that the differences between the leading singular contributions and the full NLO results are smaller for $\sqrt{s}=\unit{7}{\TeV}$ than for $\sqrt{s}=\unit{14}{\TeV}$. Here we should point out that the hadronic variable $\tau$ is about 0.025 and 0.012 for the 7 TeV and 14 TeV LHC, respectively, which is no way near the hadronic threshold $\tau \to 1$. The fact that the leading singular terms still dominate the NLO results in this case is due to the fast decreasing of the PDF as $x \to 1$. The region near the partonic threshold $z \to 1$ is therefore sampled with large weights in the convolution integral (\ref{eq:hadcs}). This also explains that the singular terms are more dominant in the $cg$ channel than in the $ug$ channel.

We now turn to the actual resummation. As discussed in the previous section, the resummation of large threshold logarithms amounts to proper choices of the two scales $\mu_h$ and $\mu_s$ for the hard function and the soft function, respectively. The hard scale $\mu_h$ should be chosen such that the hard functions evaluated at this scale have stable perturbative expansions. From Eq.~(\ref{eqs:hard_function}), we see that the natural choice is $\mu_h=m_t$, where the logarithmic terms automatically vanish. In our numerical results, we will vary $\mu_h$ around this default choice to account for the uncertainties coming from the unknown higher order corrections to the hard function.

\begin{figure}[t!]
\begin{center}
\includegraphics[width=0.45\textwidth]{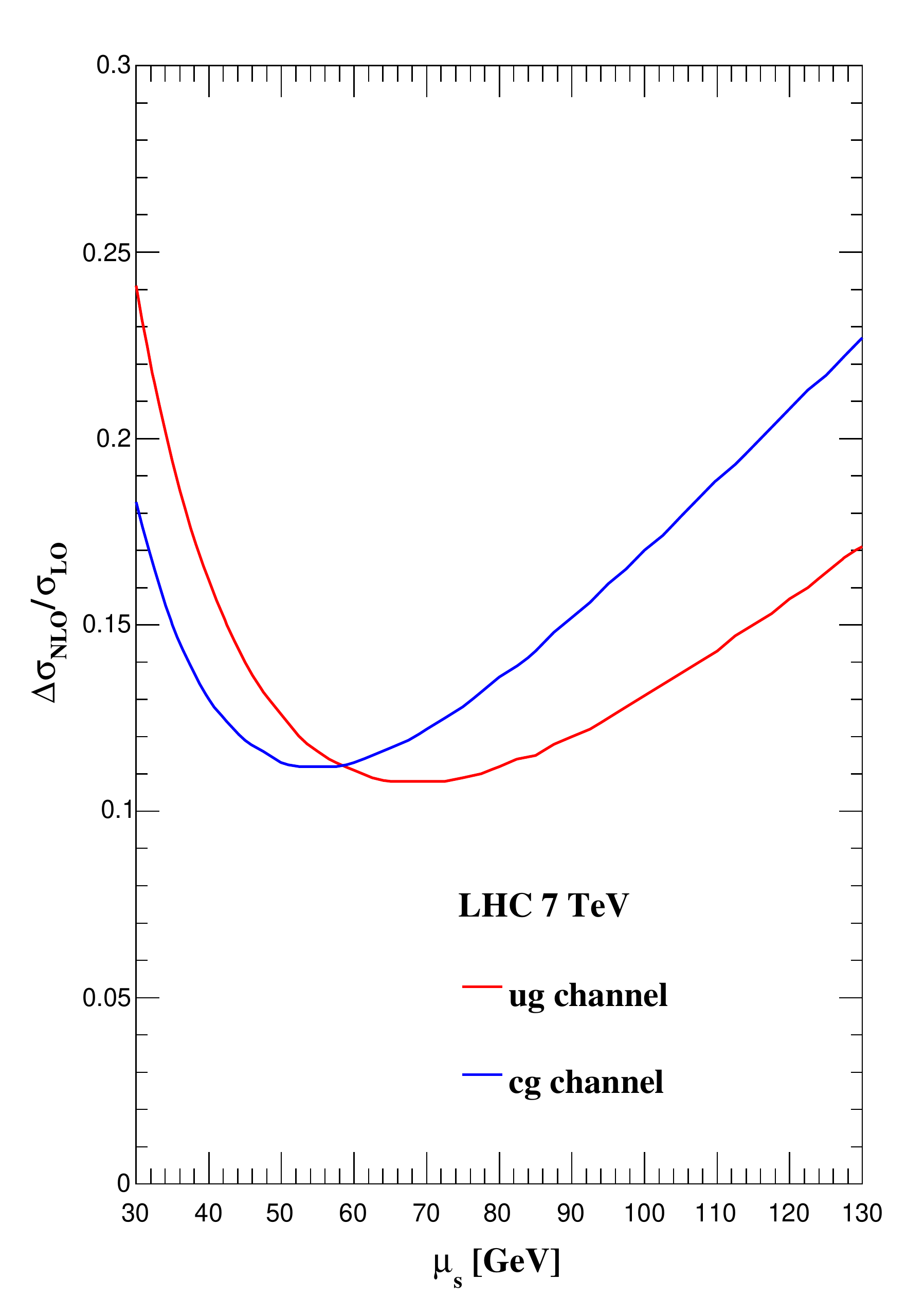}
\includegraphics[width=0.45\textwidth]{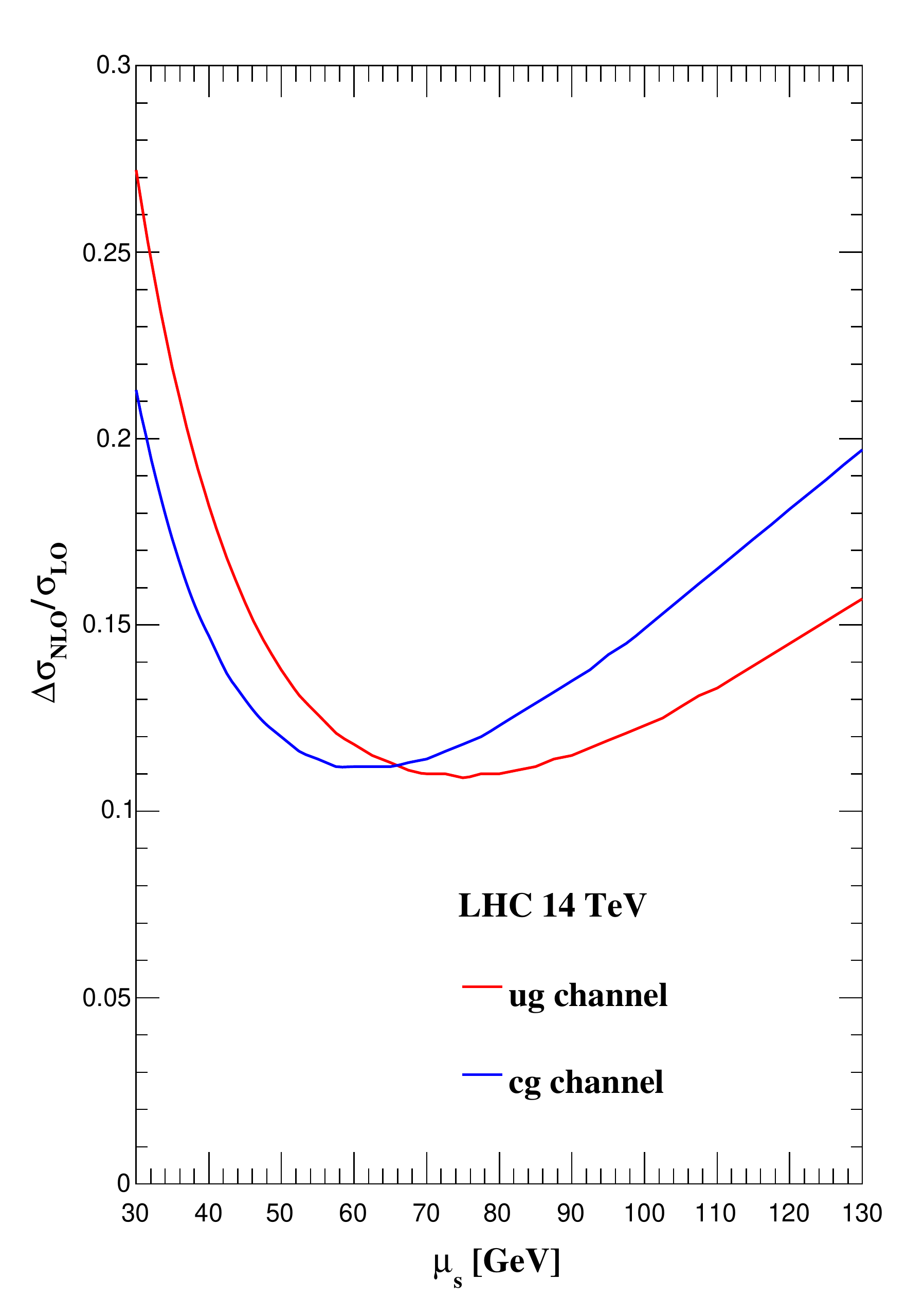}
\end{center}
\vspace{-4mm}
\caption{\label{choosemus}The contributions from the one-loop soft functions to the total cross sections in the $ug$ channel (red lines) and the $cg$ channel (blue lines).}
\end{figure}

The scenario for the soft scale is not as straightforward as the hard scale. We have discussed in the previous section that there exist different prescriptions of choosing the soft scale. Here we follow the prescription of \cite{Becher:2007ty} and choose the soft scale directly in the momentum space. Ideally, this scale should correspond to the ``average'' energy of the extra radiations in the final state. However, this average energy cannot be obtained analytically since we don't have the analytic form of the parton distributions. Therefore, we will estimate the soft scale by numerically inspecting the stability of the cross section against soft corrections. In Figure~\ref{choosemus}, we show the corrections to the cross sections from the soft function only, without the hard function and the RG evolution effects, as a function of the soft scale. We vary the soft scale from $\unit{30}{\GeV}$ to $\unit{130}{\GeV}$, and we find that the corrections are generically small for $\mu_s$ between $\unit{50}{\GeV}$ and $\unit{90}{\GeV}$. In our numerical results,
we will take the default value for $\mu_s$, denoted as $\mu_s^0$, to be $\unit{75}{\GeV}$ in the $ug$ channe and $\unit{60}{\GeV}$ in the $cg$ channel, respectively. We will also vary the soft scale around its default choices to estimate the uncertainties associated with the soft function.

\begin{figure}[t!]
\begin{center}
\includegraphics[width=0.45\textwidth]{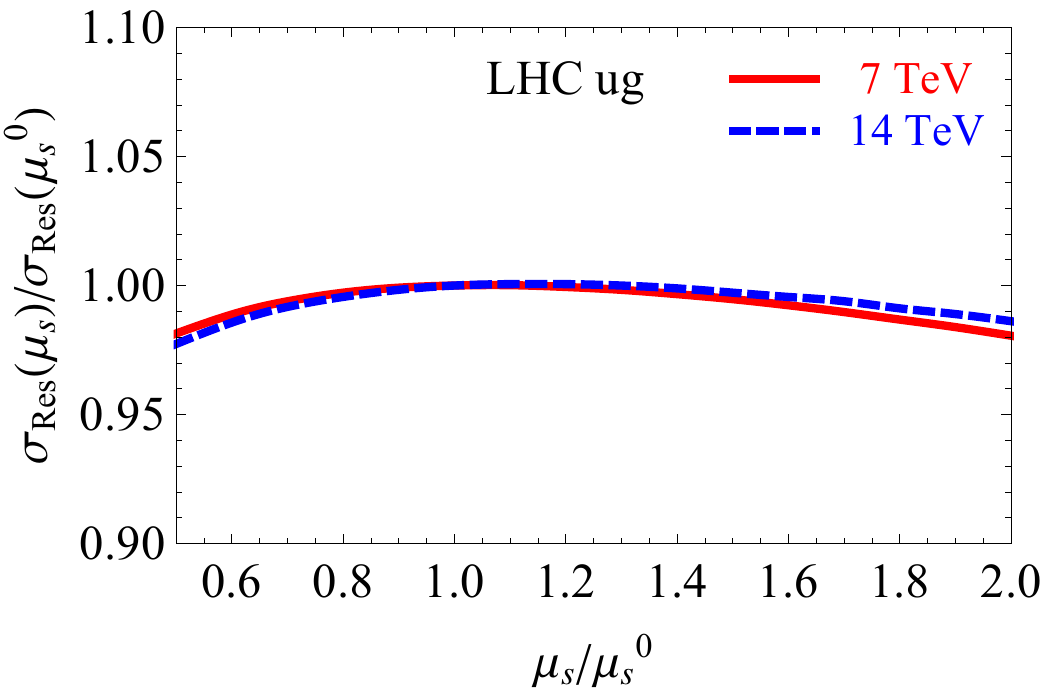}
\includegraphics[width=0.45\textwidth]{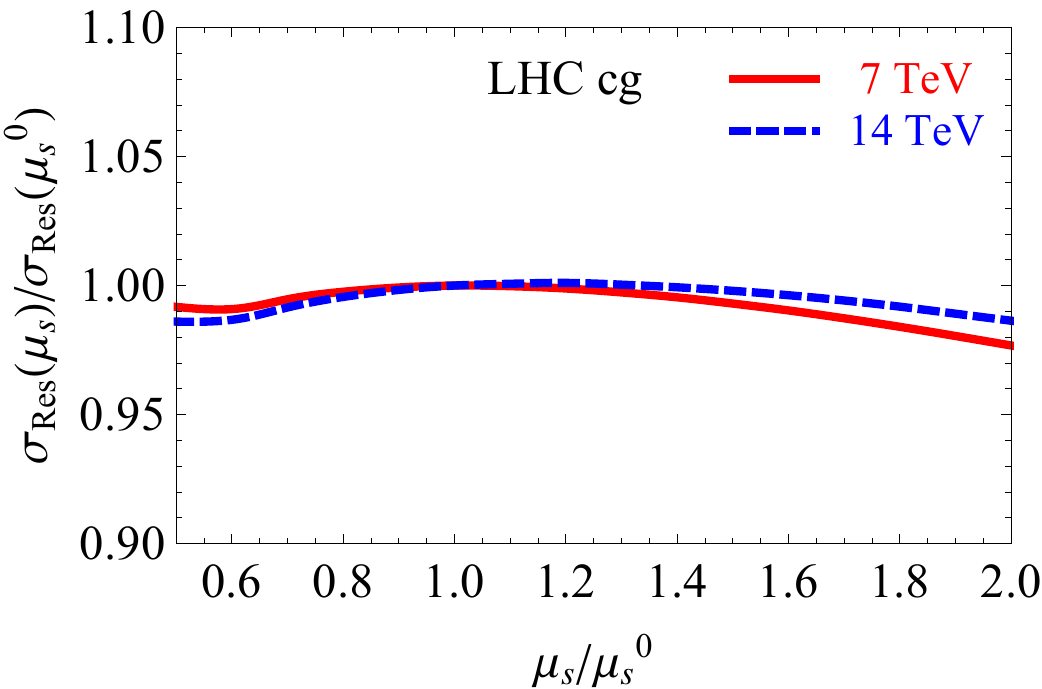}
\end{center}
\vspace{-4mm}
\caption{\label{scluncmus}
The variation of the resummed cross sections with respect to the soft scale.}
\end{figure}

\begin{figure}[t!]
\begin{center}
\includegraphics[width=0.45\textwidth]{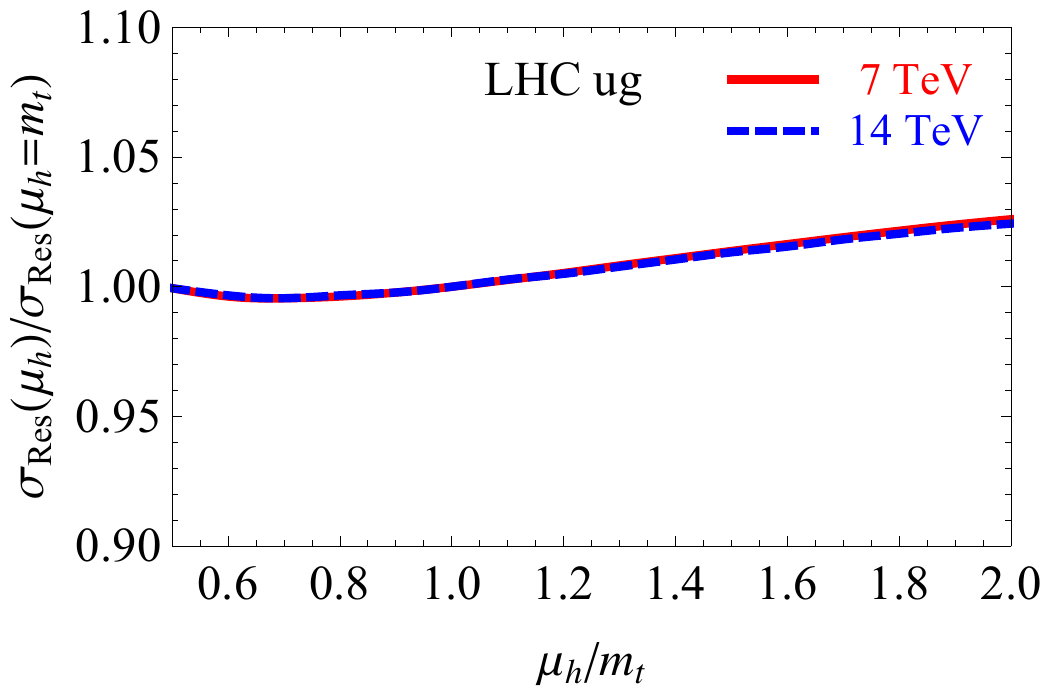}
\includegraphics[width=0.45\textwidth]{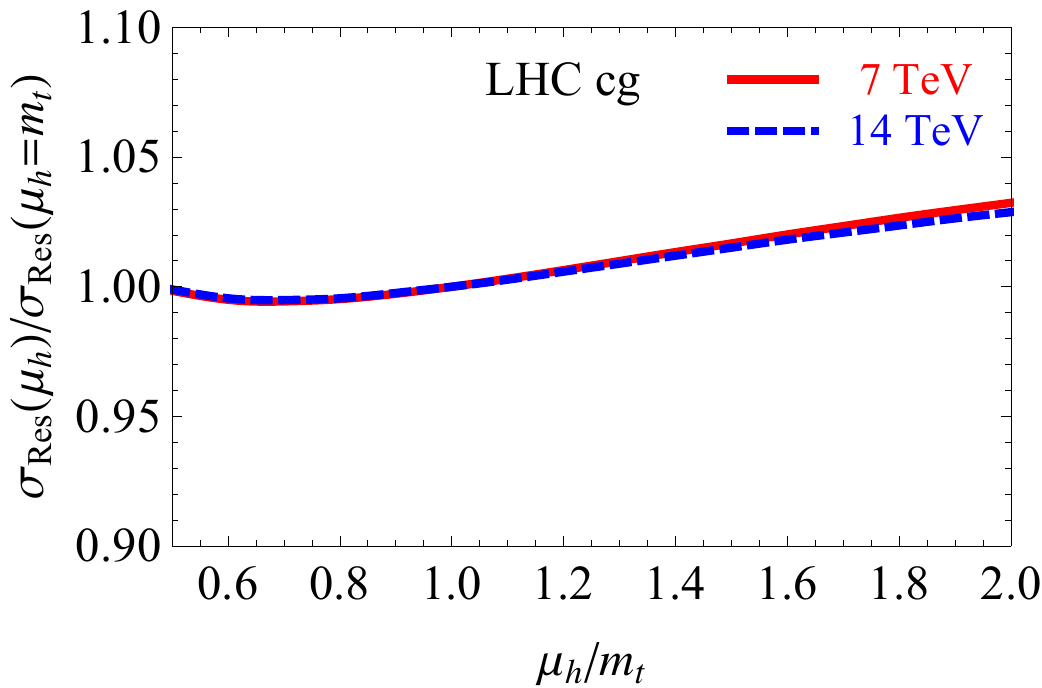}
\end{center}
\vspace{-4mm}
\caption{\label{scluncmuh}
The variation of the resummed cross sections with respect to the hard scale.}
\end{figure}

\begin{figure}[t!]
\begin{center}
\includegraphics[width=0.45\textwidth]{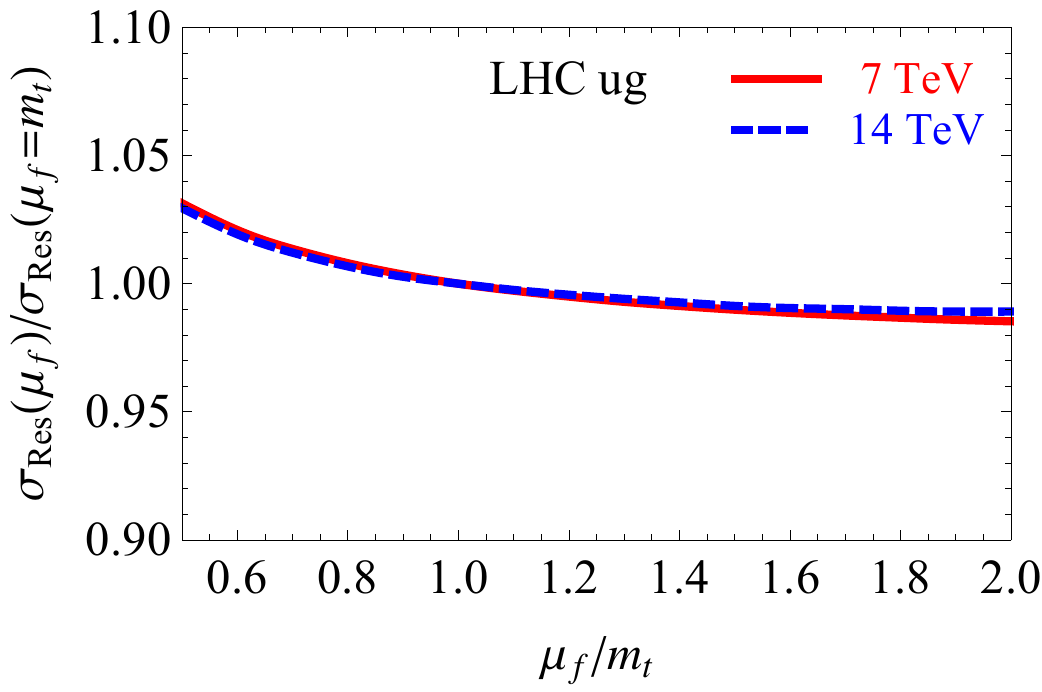}
\includegraphics[width=0.45\textwidth]{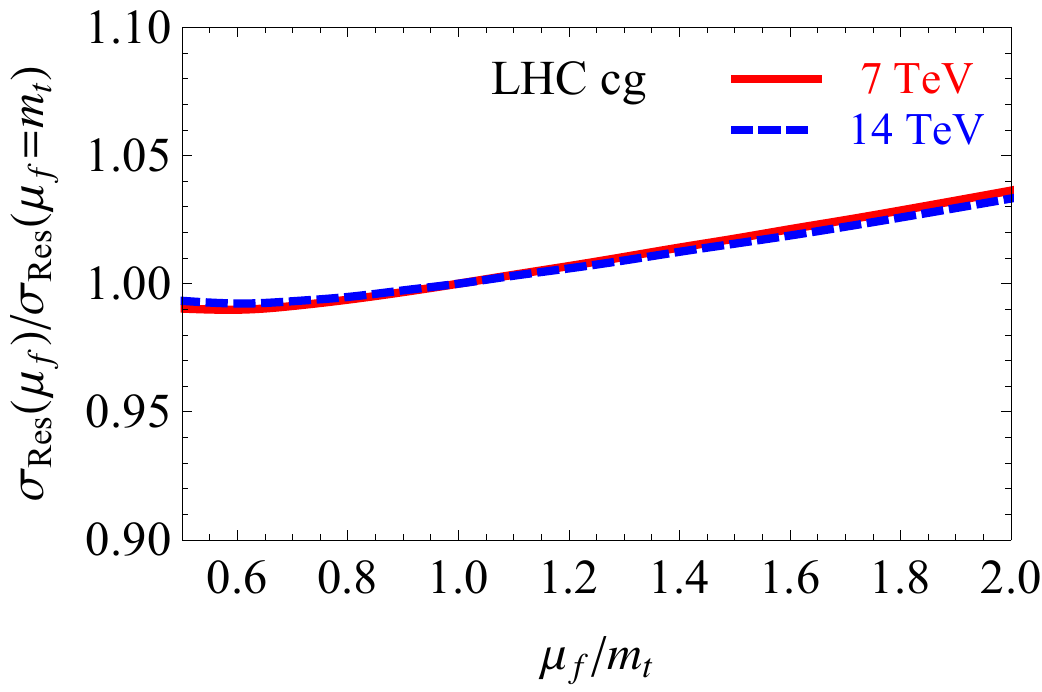}
\end{center}
\vspace{-4mm}
\caption{\label{scluncmuf}
The variation of the resummed cross sections with respect to the factorization scale.}
\end{figure}

In Figures~\ref{scluncmus}, \ref{scluncmuh} and \ref{scluncmuf}, we show the variations of the resummed cross sections with respect to the changes in the soft scale, the hard scale and the factorization scale. We see that after resummation, the cross sections are rather stable against the changes of the unphysical scales. When the scales are varied from one-half to double of the default choices, the relative changes of the cross sections are at most 4\%.


\begin{table}[t!]
\begin{center}
\begin{tabular}{lcc}
\toprule
& \multicolumn{2}{c}{LHC (\unit{7}{\TeV})}
\\
\cmidrule{2-3}
& $gu \rightarrow t$ & $gc \rightarrow t$
\\
\midrule
NLO [pb] & $6.99^{+6.3\%+3.0\%}_{-5.5\% -3.8\%}$ & $0.742^{+6.3\%+6.0\%}_{-5.6\%-5.9\%}$
\\ \addlinespace
NLO+NNLL [pb] & $7.21^{+4.0\%+3.1\%}_{-2.6\%-3.9\%}$ & $0.790^{+4.1\%+6.0\%}_{-2.5\%-5.9\%}$
\\
\specialrule{\heavyrulewidth}{\aboverulesep}{\belowrulesep}
& \multicolumn{2}{c}{LHC (\unit{8}{\TeV})}
\\
\cmidrule{2-3}
& $gu \rightarrow t$ & $gc \rightarrow t$
\\
\midrule
NLO [pb] & $8.44^{+6.4\%+3.1\%}_{-5.5\%-4.0\%}$ & $0.978^{+6.4\%+5.7\%}_{-5.7\%-5.7\%}$
\\ \addlinespace
NLO+NNLL [pb] & $8.69^{+4.0\%+3.1\%}_{-2.6\%-4.0\%}$ & $1.04^{+4.1\%+5.7\%}_{-2.4\%-5.7\%}$
\\
\specialrule{\heavyrulewidth}{\aboverulesep}{\belowrulesep}
& \multicolumn{2}{c}{LHC (\unit{13}{\TeV})}
\\
\cmidrule{2-3}
& $gu \rightarrow t$ & $gc \rightarrow t$
\\
\hline
NLO [pb] & $16.1^{+6.5\%+3.3\%}_{-5.7\%-4.5\%}$ & $2.47^{+6.7\%+5.0\%}_{-6.2\%-5.4\%}$
\\ \addlinespace
NLO+NNLL [pb] & $16.5^{+3.8\%+3.3\%}_{-2.7\%-4.5\%}$ & $2.58^{+3.8\%+5.1\%  }_{-2.7\%-5.5\%}$
\\
\specialrule{\heavyrulewidth}{\aboverulesep}{\belowrulesep}
& \multicolumn{2}{c}{LHC (\unit{14}{\TeV})}
\\
\cmidrule{2-3}
& $gu \rightarrow t$ & $gc \rightarrow t$
\\
\midrule
NLO [pb] & $17.7^{+6.6\%+3.3\%}_{-5.8\%-4.5\%}$ & $2.82^{+6.8\%+5.0\%}_{-6.3\%-5.5\%}$
\\ \addlinespace
NLO+NNLL [pb] & $18.1^{+3.8\%+3.4\%}_{-2.7\%-4.6\%}$ & $2.95^{+3.8\%+5.1\%}_{-2.8\%-5.5\%}$
\\
\bottomrule
\end{tabular}
\caption{The predictions for the cross sections at the LHC with $\sqrt{s} = 7, 8, 13, \unit{14}{\TeV}$. The first uncertainties are estimated by varying unphysical scales in the calculation, while the second uncertainties come from the experimental error of the PDFs and the strong coupling constant $\alpha_s$. The anomalous couplings are taken to be $\kappa_{tqg}/\Lambda = \unit{0.01}{\TinveV}$.}
\label{tab:tot}
\end{center}
\end{table}

We now give our final predictions for the total cross sections in Table~\ref{tab:tot}. The NLO and NLO+NNLL cross sections are shown for the LHC with center-of-mass energies 7, 8, 13 and \unit{14}{\TeV}. In the table two kinds of uncertainties are shown. The first uncertainties are due to unknown higher order perturbative corrections, and are estimated by varying the unphysical scales in our calculation. More precisely, for the NLO cross sections, the renormalization scale $\mu_r$ and the factorization scale $\mu_f$ are varied separately between $m_t/2$ and $2m_t$. The resulting uncertainties in the cross sections are then added in quadrature. For the NLO+NNLL resummed cross sections, there are three unphysical scales: the hard scale $\mu_h$, the soft scale $\mu_s$, as well as the factorization scale $\mu_f$. These are again separately varied between one-half and double of their default values, and the final combined uncertainties are obtained by adding in quadrature. The second uncertainties come from the experimental error on the PDFs and the strong coupling constant $\alpha_s$, and are estimated using the CT10 PDF sets according to \cite{Lai:2010vv}. From the numbers in the table, we see that the soft gluon effects lead to a moderate increase of the cross sections by a few percents. The scale uncertainties are reduced when going from the NLO results to the NLO+NNLL resummed results. This can be expected due to the higher-order corrections included in the resummed formula. On the other hand, the PDF+$\alpha_s$ uncertainties, being an experimental effect, are largely the same between NLO and NLO+NNLL results.

\section{Conclusion}
\label{sec:conclusion}

In conclusion, we have upgraded our predictions of the total cross sections for direct top quark production. Our new predictions are based on an NLO+NNLL resummed formula, incorporating the prescription of choosing the soft scale directly in the momentum space. The soft gluon effects lead to an increase of the total cross sections by a few percents, and stabilize the variation of the cross sections against the changes in the unphysical scales.

\section*{Acknowledgments}
We would like to thank Ding Yu Shao for collaboration at the early stage of this work.
We also appreciate Lorenzo Basso for pointing out the misplacement of two plots in figure \ref{fig:singular}.
This work was supported in part by the National Natural Science Foundation of China, under Grants No. 11375013 and No. 11135003,
and by the U.S. DOE Early Career Research Award DE-SC0003870 by Lightner-Sams Foundation.
The research of J.W. has been supported by
the Cluster of Excellence {\it Precision Physics, Fundamental Interactions and Structure of Matter} (PRISMA-EXC 1098).

\appendix

\section{Anomalous dimensions}
\label{app:a}

In this appendix we collect various anomalous dimensions and relevant expressions used in our resummation formula. First of all, we have two coupling constants in our process, the strong coupling constant $\alpha_s$ and the anomalous flavor-changing coupling constant $\kappa_{tqg}$. The strong coupling constant satisfies the well-known RG equation with the $\beta$-function
\begin{align}
 \beta(\alpha_s) = -2 \alpha_s \left[ \beta_0 \frac{\alpha_s}{4\pi} + \beta_1 \left(\frac{\alpha_s}{4\pi} \right)^2 + \beta_2 \left(\frac{\alpha_s}{4\pi} \right)^3 + \cdots \right] ,
\end{align}
where the coefficients read
\begin{align}
 \beta_0 &= \frac{11}{3} C_A - \frac{4}{3} T_F n_f \, , \nonumber
 \\
 \beta_1 &= \frac{34}{3}C^2_A - \frac{20}{3} C_A T_F n_f - 4C_F T_F n_f \, , \nonumber
 \\
 \beta_2 &= \frac{2857}{54} C^3_A + \left( 2 C^2_F - \frac{205}{9} C_F C_A - \frac{1415}{27} C^2_A \right) T_F n_f + \left( \frac{44}{9}C_F + \frac{158}{27} C_A \right) T^2_F n^2_f \, .
\end{align}
The anomalous coupling also satisfies a RG equation
\begin{align}\label{eqs:rge}
  \frac{d\kappa(\mu)}{d\ln\mu} = \gamma^\kappa(\alpha_s) \, \kappa(\mu) \, .
\end{align}
The one-loop anomalous dimension has been obtained in \cite{Yang:2006gs}, while the two-loop anomalous dimension can be extracted from the results in \cite{Buchalla:1995vs}. They read
\begin{align}
  \gamma^\kappa_0 &= \frac{4}{3} \, , \nonumber
  \\
  \gamma^\kappa_1 &= \frac{244}{9} + \frac{59}{27} \, n_f \, .
\end{align}
Here and below, all coefficients of anomalous dimensions are in terms of an expansion in $\alpha_s/4\pi$.

The coefficients of the cusp anomalous dimension are
\begin{align}
  \gamma^{\text{cusp}}_0 &= 4 \, , \nonumber
  \\
  \gamma^{\text{cusp}}_1 &= 4 \left[ \left( \frac{67}{9} - \frac{\pi^2}{3} \right) C_A - \frac{20}{9} T_F n_f \right] , \nonumber
  \\
  \gamma^{\text{cusp}}_2 &= 4 \left[ C^2_A \left(\frac{245}{6} - \frac{134}{27}\pi^2 + \frac{11}{45}\pi^4 + \frac{22}{3}\zeta_3 \right) + C_A T_F n_f \left(-\frac{418}{27} + \frac{40}{27}\pi^2 - \frac{56}{3}\zeta_3 \right) \right. \nonumber
  \\
  &\left. + C_F T_F n_f \left( -\frac{55}{3} + 16 \zeta_3 \right) - \frac{16}{27} T^2_F n^2_f \right] ,
\end{align}
and the other relevant anomalous dimensions are given by
\begin{align*}
  \gamma_0^g &= - \beta_0 = - \frac{11}{3}\,C_A + \frac43\,T_F n_f \, ,
  \\
  \gamma_1^g &= C_A^2 \left( -\frac{692}{27} + \frac{11\pi^2}{18} + 2\zeta_3 \right) + C_A T_F n_f \left( \frac{256}{27} - \frac{2\pi^2}{9} \right) + 4 C_F T_F n_f \, ,
  \\
  \gamma^q_0 &= -3C_F \, ,
  \\
  \gamma^q_1 &= C^2_F \left( -\frac{3}{2} + 2\pi^2 - 24 \zeta_3 \right) + C_F C_A \left( - \frac{961}{54} - \frac{11}{6}\pi^2 + 26\zeta_3 \right) + C_F T_F n_f \left( \frac{130}{27} + \frac{2}{3}\pi^2 \right) ,
  \\
  \gamma^Q_0 &= -2C_F ,
  \\
  \gamma^Q_1 &= C_F C_A\left(\frac{2}{3}\pi^2-\frac{98}{9}-4\zeta_3 \right)+\frac{40}{9}C_F T_F n_f,
  \\
  \gamma^{\phi_q}_0 &= 3C_F \, ,
  \\
  \gamma^{\phi_q}_1 &= C^2_F \left( \frac{3}{2} - 2 \pi^2 + 24 \zeta_3 \right) + C_F C_A \left( \frac{17}{6} + \frac{22}{9} \pi^2 - 12 \zeta_3 \right) - C_F T_F n_f \left( \frac{2}{3} + \frac{8}{9} \pi^2 \right) ,
  \\
  \gamma^{\phi_g}_0 &= \beta_0 \, ,
  \\
  \gamma^{\phi_g}_1 &= 4C^2_A \left( \frac{8}{3} + 3 \zeta_3 \right) - \frac{16}{3} C_A T_F n_f - 4C_F T_F n_f \, ,
\end{align*}

The anomalous dimensions for the hard and soft functions can be obtained from the above anomalous dimensions through the following relations:
\begin{align}
  \gamma^H(\alpha_s) &= \gamma^g(\alpha_s) + \gamma^q(\alpha_s) + \gamma^Q(\alpha_s) \, , \nonumber
  \\
  \gamma^S(\alpha_s) &= \gamma^{H}(\alpha_s) + \gamma^{\phi_g}(\alpha_s) + \gamma^{\phi_q}(\alpha_s) \, .
\end{align}
The explicit expressions for the first two expansion coefficients of $\gamma^H(\alpha_s)$ are given by
\begin{align}
  \gamma^H_0 &= -\frac{53}{3} + \frac{2n_f}{3} \, , \nonumber
  \\
  \gamma^H_1 &= -\frac{9398}{27} + \frac{79\pi^2}{18} + \frac{190\zeta_3}{3} + n_f \left(
    \frac{1868}{81} + \frac{\pi^2}{9} \right) \, .
\end{align}


\begin{thebibliography}{99}

%\cite{Aad:2012tfa}
\bibitem{Aad:2012tfa}
  G.~Aad {\it et al.}  [ATLAS Collaboration],
  %``Observation of a new particle in the search for the Standard Model Higgs boson with the ATLAS detector at the LHC,''
  Phys.\ Lett.\ B {\bf 716}, 1 (2012)
  [arXiv:1207.7214 [hep-ex]].
  %%CITATION = ARXIV:1207.7214;%%
  %3265 citations counted in INSPIRE as of 16 Sep 2014


%\cite{Chatrchyan:2012ufa}
\bibitem{Chatrchyan:2012ufa}
  S.~Chatrchyan {\it et al.}  [CMS Collaboration],
  %``Observation of a new boson at a mass of 125 GeV with the CMS experiment at the LHC,''
  Phys.\ Lett.\ B {\bf 716}, 30 (2012)
  [arXiv:1207.7235 [hep-ex]].
  %%CITATION = ARXIV:1207.7235;%%
  %3210 citations counted in INSPIRE as of 16 Sep 2014


%\cite{ATLAS-CONF-2013-102}
\bibitem{ATLAS-CONF-2013-102}
  ATLAS collaboration,
  ``Combination of ATLAS and CMS results on the mass of the top-quark using up to 4.9 fb$^{-1}$ of $\sqrt{s}=7$ TeV LHC data,''
  ATLAS-CONF-2013-102.
  %%CITATION = ATLAS-CONF-2013-102;%%
  %14 citations counted in INSPIRE as of 16 Sep 2014


%\cite{CMS-PAS-TOP-13-005}
\bibitem{CMS-PAS-TOP-13-005}
  CMS Collaboration,
  ``Combination of ATLAS and CMS results on the mass of the top quark using up to 4.9 inverse femtobarns of data,''
  CMS-PAS-TOP-13-005.
  %%CITATION = CMS-PAS-TOP-13-005;%%
  %9 citations counted in INSPIRE as of 16 Sep 2014


%\cite{Tevatron:2014cka}
\bibitem{Tevatron:2014cka}
  Tevatron Electroweak Working Group [CDF and D0 Collaborations],
  %``Combination of CDF and D0 results on the mass of the top quark using up to 9.7 fb$^{-1}$ at the Tevatron,''
  arXiv:1407.2682 [hep-ex].
  %%CITATION = ARXIV:1407.2682;%%
  %5 citations counted in INSPIRE as of 16 Sep 2014


%\cite{ATLAS:2014wva}
\bibitem{ATLAS:2014wva}
  ATLAS and CDF and CMS and D0 Collaborations,
  %``First combination of Tevatron and LHC measurements of the top-quark mass,''
  arXiv:1403.4427 [hep-ex].
  %%CITATION = ARXIV:1403.4427;%%
  %68 citations counted in INSPIRE as of 16 Sep 2014


%\cite{Baernreuther:2012ws}
\bibitem{Baernreuther:2012ws}
  P.~B?rnreuther, M.~Czakon and A.~Mitov,
  %``Percent Level Precision Physics at the Tevatron: First Genuine NNLO QCD Corrections to $q \bar{q} \to t \bar{t} + X$,''
  Phys.\ Rev.\ Lett.\  {\bf 109}, 132001 (2012)
  [arXiv:1204.5201 [hep-ph]].
  %%CITATION = ARXIV:1204.5201;%%
  %176 citations counted in INSPIRE as of 16 Sep 2014


%\cite{Czakon:2012zr}
\bibitem{Czakon:2012zr}
  M.~Czakon and A.~Mitov,
  %``NNLO corrections to top-pair production at hadron colliders: the all-fermionic scattering channels,''
  JHEP {\bf 1212}, 054 (2012)
  [arXiv:1207.0236 [hep-ph]].
  %%CITATION = ARXIV:1207.0236;%%
  %113 citations counted in INSPIRE as of 16 Sep 2014


%\cite{Czakon:2012pz}
\bibitem{Czakon:2012pz}
  M.~Czakon and A.~Mitov,
  %``NNLO corrections to top pair production at hadron colliders: the quark-gluon reaction,''
  JHEP {\bf 1301}, 080 (2013)
  [arXiv:1210.6832 [hep-ph]].
  %%CITATION = ARXIV:1210.6832;%%
  %107 citations counted in INSPIRE as of 16 Sep 2014


%\cite{Czakon:2013goa}
\bibitem{Czakon:2013goa}
  M.~Czakon, P.~Fiedler and A.~Mitov,
  %``Total Top-Quark Pair-Production Cross Section at Hadron Colliders Through $O(α\frac{4}{S})$,''
  Phys.\ Rev.\ Lett.\  {\bf 110}, no. 25, 252004 (2013)
  [arXiv:1303.6254 [hep-ph]].
  %%CITATION = ARXIV:1303.6254;%%
  %245 citations counted in INSPIRE as of 16 Sep 2014


%\cite{Ahrens:2010zv}
\bibitem{Ahrens:2010zv}
  V.~Ahrens, A.~Ferroglia, M.~Neubert, B.~D.~Pecjak and L.~L.~Yang,
  %``Renormalization-Group Improved Predictions for Top-Quark Pair Production at Hadron Colliders,''
  JHEP {\bf 1009}, 097 (2010)
  [arXiv:1003.5827 [hep-ph]].
  %%CITATION = ARXIV:1003.5827;%%
  %204 citations counted in INSPIRE as of 16 Sep 2014


%\cite{Ahrens:2011mw}
\bibitem{Ahrens:2011mw}
  V.~Ahrens, A.~Ferroglia, M.~Neubert, B.~D.~Pecjak and L.~L.~Yang,
  %``RG-improved single-particle inclusive cross sections and forward-backward asymmetry in $t\bar t$ production at hadron colliders,''
  JHEP {\bf 1109}, 070 (2011)
  [arXiv:1103.0550 [hep-ph]].
  %%CITATION = ARXIV:1103.0550;%%
  %74 citations counted in INSPIRE as of 16 Sep 2014

%\cite{Zhu:2012ts}
\bibitem{Zhu:2012ts}
  H.~X.~Zhu, C.~S.~Li, H.~T.~Li, D.~Y.~Shao and L.~L.~Yang,
  %``Transverse-momentum resummation for top-quark pairs at hadron colliders,''
  Phys.\ Rev.\ Lett.\  {\bf 110}, 082001 (2013)
  [arXiv:1208.5774 [hep-ph]].
  %%CITATION = ARXIV:1208.5774;%%
  %19 citations counted in INSPIRE as of 22 Sep 2014

%\cite{Broggio:2014yca}
\bibitem{Broggio:2014yca}
  A.~Broggio, A.~S.~Papanastasiou and A.~Signer,
  %``Renormalization-group improved fully differential cross sections for top pair production,''
  arXiv:1407.2532 [hep-ph].
  %%CITATION = ARXIV:1407.2532;%%
  %4 citations counted in INSPIRE as of 16 Sep 2014

%\cite{Kidonakis:2010tc}
\bibitem{Kidonakis:2010tc}
  N.~Kidonakis,
  %``NNLL resummation for s-channel single top quark production,''
  Phys.\ Rev.\ D {\bf 81}, 054028 (2010)
  [arXiv:1001.5034 [hep-ph]].
  %%CITATION = ARXIV:1001.5034;%%
  %278 citations counted in INSPIRE as of 17 Sep 2014

%\cite{Zhu:2010mr}
\bibitem{Zhu:2010mr}
  H.~X.~Zhu, C.~S.~Li, J.~Wang and J.~J.~Zhang,
  %``Factorization and resummation of s-channel single top quark production,''
  JHEP {\bf 1102}, 099 (2011)
  [arXiv:1006.0681 [hep-ph]].
  %%CITATION = ARXIV:1006.0681;%%
  %22 citations counted in INSPIRE as of 16 Sep 2014


%\cite{Wang:2012dc}
\bibitem{Wang:2012dc}
  J.~Wang, C.~S.~Li and H.~X.~Zhu,
  %``Resummation prediction on top quark transverse momentum distribution at large $p_T$,''
  Phys.\ Rev.\ D {\bf 87},  034030 (2013)
  [arXiv:1210.7698 [hep-ph]].
  %%CITATION = ARXIV:1210.7698;%%
  %5 citations counted in INSPIRE as of 16 Sep 2014

%\cite{Kidonakis:2011wy}
\bibitem{Kidonakis:2011wy}
  N.~Kidonakis,
  %``Next-to-next-to-leading-order collinear and soft gluon corrections for t-channel single top quark production,''
  Phys.\ Rev.\ D {\bf 83}, 091503 (2011)
  [arXiv:1103.2792 [hep-ph]].
  %%CITATION = ARXIV:1103.2792;%%
  %299 citations counted in INSPIRE as of 17 Sep 2014

%\cite{Gao:2012ja}
\bibitem{Gao:2012ja}
  J.~Gao, C.~S.~Li and H.~X.~Zhu,
  %``Top Quark Decay at Next-to-Next-to Leading Order in QCD,''
  Phys.\ Rev.\ Lett.\  {\bf 110}, 042001 (2013)
  [arXiv:1210.2808 [hep-ph]].
  %%CITATION = ARXIV:1210.2808;%%
  %28 citations counted in INSPIRE as of 16 Sep 2014


%\cite{Brucherseifer:2013iv}
\bibitem{Brucherseifer:2013iv}
  M.~Brucherseifer, F.~Caola and K.~Melnikov,
  %``$\mathcal O(\alpha_s^2)$ corrections to fully-differential top quark decays,''
  JHEP {\bf 1304}, 059 (2013)
  [arXiv:1301.7133 [hep-ph]].
  %%CITATION = ARXIV:1301.7133;%%
  %20 citations counted in INSPIRE as of 16 Sep 2014


%\cite{Aad:2012gd}
\bibitem{Aad:2012gd}
  G.~Aad {\it et al.}  [ATLAS Collaboration],
  %``Search for FCNC single top-quark production at $\sqrt{s}=7$ TeV with the ATLAS detector,''
  Phys.\ Lett.\ B {\bf 712}, 351 (2012)
  [arXiv:1203.0529 [hep-ex]].
  %%CITATION = ARXIV:1203.0529;%%
  %56 citations counted in INSPIRE as of 16 Sep 2014


%\cite{Aad:2012ij}
\bibitem{Aad:2012ij}
  G.~Aad {\it et al.}  [ATLAS Collaboration],
  %``A search for flavour changing neutral currents in top-quark decays in $pp$ collision data collected with the ATLAS detector at $\sqrt{s}=7$ TeV,''
  JHEP {\bf 1209}, 139 (2012)
  [arXiv:1206.0257 [hep-ex]].
  %%CITATION = ARXIV:1206.0257;%%
  %43 citations counted in INSPIRE as of 16 Sep 2014


%\cite{Chatrchyan:2012hqa}
\bibitem{Chatrchyan:2012hqa}
  S.~Chatrchyan {\it et al.}  [CMS Collaboration],
  %``Search for flavor changing neutral currents in top quark decays in pp collisions at 7 TeV,''
  Phys.\ Lett.\ B {\bf 718}, 1252 (2013)
  [arXiv:1208.0957 [hep-ex]].
  %%CITATION = ARXIV:1208.0957;%%
  %35 citations counted in INSPIRE as of 16 Sep 2014


%\cite{Chatrchyan:2013nwa}
\bibitem{Chatrchyan:2013nwa}
  S.~Chatrchyan {\it et al.}  [CMS Collaboration],
  %``Search for flavor-changing neutral currents in top-quark decays t to Zq in pp collisions at sqrt(s) = 8 TeV,''
  Phys.\ Rev.\ Lett.\  {\bf 112}, 171802 (2014)
  [arXiv:1312.4194 [hep-ex]].
  %%CITATION = ARXIV:1312.4194;%%
  %10 citations counted in INSPIRE as of 16 Sep 2014


%\cite{CMS-PAS-TOP-12-037}
\bibitem{CMS-PAS-TOP-12-037}
  CMS Collaboration [CMS Collaboration],
  ``Search for flavor changing neutral currents in top quark decays
in pp collisions at 8 TeV,''
  CMS-PAS-TOP-12-037.
  %%CITATION = CMS-PAS-TOP-12-037;%%
  %9 citations counted in INSPIRE as of 16 Sep 2014


%\cite{ATLAS-CONF-2013-063}
\bibitem{ATLAS-CONF-2013-063}
  ATLAS collaboration,
  ``Search for single top-quark production via FCNC in strong interaction in $\sqrt{s}=8\,\,\mathrm{TeV}$ ATLAS data,''
  ATLAS-CONF-2013-063.
  %%CITATION = ATLAS-CONF-2013-063;%%
  %10 citations counted in INSPIRE as of 16 Sep 2014


%\cite{CMS-PAS-HIG-13-034}
\bibitem{CMS-PAS-HIG-13-034}
  CMS Collaboration,
  ``Combined multilepton and diphoton limit on $t \to cH$,''
  CMS-PAS-HIG-13-034.
  %%CITATION = CMS-PAS-HIG-13-034;%%
  %9 citations counted in INSPIRE as of 16 Sep 2014


%\cite{CMS-PAS-TOP-14-003}
\bibitem{CMS-PAS-TOP-14-003}
  CMS Collaboration,
  ``Search for anomalous single top quark production in
association with a photon,''
  CMS-PAS-TOP-14-003.
  %%CITATION = CMS-PAS-TOP-14-003;%%
  %2 citations counted in INSPIRE as of 16 Sep 2014


%\cite{CMS-PAS-TOP-14-007}
\bibitem{CMS-PAS-TOP-14-007}
  CMS Collaboration,
  ``Search for anomalous Wtb couplings and top FCNC in t-channel single-top-quark events,''
  CMS-PAS-TOP-14-007.
  %%CITATION = CMS-PAS-TOP-14-007;%%
  %1 citations counted in INSPIRE as of 16 Sep 2014


%\cite{Liu:2005dp}
\bibitem{Liu:2005dp}
  J.~J.~Liu, C.~S.~Li, L.~L.~Yang and L.~G.~Jin,
  %``Next-to-leading order QCD corrections to the direct top quark production via model-independent FCNC couplings at hadron colliders,''
  Phys.\ Rev.\ D {\bf 72}, 074018 (2005)
  [hep-ph/0508016].
  %%CITATION = HEP-PH/0508016;%%
  %34 citations counted in INSPIRE as of 16 Sep 2014


%\cite{Gao:2011fx}
\bibitem{Gao:2011fx}
  J.~Gao, C.~S.~Li, L.~L.~Yang and H.~Zhang,
  %``Search for anomalous top quark production at the early LHC,''
  Phys.\ Rev.\ Lett.\  {\bf 107}, 092002 (2011)
  [arXiv:1104.4945 [hep-ph]].
  %%CITATION = ARXIV:1104.4945;%%
  %17 citations counted in INSPIRE as of 16 Sep 2014

%\cite{Glashow:1970gm}
\bibitem{Glashow:1970gm}
  S.~L.~Glashow, J.~Iliopoulos and L.~Maiani,
  %``Weak Interactions with Lepton-Hadron Symmetry,''
  Phys.\ Rev.\ D {\bf 2}, 1285 (1970).
  %%CITATION = PHRVA,D2,1285;%%
  %4754 citations counted in INSPIRE as of 13 Nov 2014

%\cite{AguilarSaavedra:2004wm}
\bibitem{AguilarSaavedra:2004wm}
  J.~A.~Aguilar-Saavedra,
  %``Top flavor-changing neutral interactions: Theoretical expectations and experimental detection,''
  Acta Phys.\ Polon.\ B {\bf 35}, 2695 (2004)
  [hep-ph/0409342].
  %%CITATION = HEP-PH/0409342;%%
  %180 citations counted in INSPIRE as of 13 Nov 2014

%\cite{Zhang:2008yn}
\bibitem{Zhang:2008yn}
  J.~J.~Zhang, C.~S.~Li, J.~Gao, H.~Zhang, Z.~Li, C.-P.~Yuan and T.~C.~Yuan,
  %``Next-to-leading order QCD corrections to the top quark decay via model-independent FCNC couplings,''
  Phys.\ Rev.\ Lett.\  {\bf 102}, 072001 (2009)
  [arXiv:0810.3889 [hep-ph]].
  %%CITATION = ARXIV:0810.3889;%%
  %42 citations counted in INSPIRE as of 16 Sep 2014


%\cite{Zhang:2010bm}
\bibitem{Zhang:2010bm}
  J.~J.~Zhang, C.~S.~Li, J.~Gao, H.~X.~Zhu, C.-P.~Yuan and T.~C.~Yuan,
  %``Next-to-leading order QCD corrections to the top quark decay via the Flavor-Changing Neutral-Current operators with mixing effects,''
  Phys.\ Rev.\ D {\bf 82}, 073005 (2010)
  [arXiv:1004.0898 [hep-ph]].
  %%CITATION = ARXIV:1004.0898;%%
  %15 citations counted in INSPIRE as of 16 Sep 2014


%\cite{Drobnak:2010wh}
\bibitem{Drobnak:2010wh}
  J.~Drobnak, S.~Fajfer and J.~F.~Kamenik,
  %``Flavor Changing Neutral Coupling Mediated Radiative Top Quark Decays at Next-to-Leading Order in QCD,''
  Phys.\ Rev.\ Lett.\  {\bf 104}, 252001 (2010)
  [arXiv:1004.0620 [hep-ph]].
  %%CITATION = ARXIV:1004.0620;%%
  %24 citations counted in INSPIRE as of 16 Sep 2014


%\cite{Drobnak:2010by}
\bibitem{Drobnak:2010by}
  J.~Drobnak, S.~Fajfer and J.~F.~Kamenik,
  %``QCD Corrections to Flavor Changing Neutral Coupling Mediated Rare Top Quark Decays,''
  Phys.\ Rev.\ D {\bf 82}, 073016 (2010)
  [arXiv:1007.2551 [hep-ph]].
  %%CITATION = ARXIV:1007.2551;%%
  %14 citations counted in INSPIRE as of 16 Sep 2014


%\cite{Yang:2006gs}
\bibitem{Yang:2006gs}
  L.~L.~Yang, C.~S.~Li, Y.~Gao and J.~J.~Liu,
  %``Threshold resummation effects in direct top quark production at hadron colliders,''
  Phys.\ Rev.\ D {\bf 73}, 074017 (2006)
  [hep-ph/0601180].
  %%CITATION = HEP-PH/0601180;%%
  %23 citations counted in INSPIRE as of 16 Sep 2014

%\cite{Kidonakis:2014dua}
\bibitem{Kidonakis:2014dua}
  N.~Kidonakis and E.~Martin,
  %``Soft-Gluon Corrections in FCNC Top-Quark Production via Anomalous Gluon Couplings,''
  arXiv:1404.7488 [hep-ph].
  %%CITATION = ARXIV:1404.7488;%%

%\cite{Catani:1996yz}
\bibitem{Catani:1996yz}
  S.~Catani, M.~L.~Mangano, P.~Nason and L.~Trentadue,
  %``The Resummation of soft gluons in hadronic collisions,''
  Nucl.\ Phys.\ B {\bf 478}, 273 (1996)
  [hep-ph/9604351].
  %%CITATION = HEP-PH/9604351;%%
  %340 citations counted in INSPIRE as of 16 Sep 2014


%\cite{Abbate:2007qv}
\bibitem{Abbate:2007qv}
  R.~Abbate, S.~Forte and G.~Ridolfi,
  %``A New prescription for soft gluon resummation,''
  Phys.\ Lett.\ B {\bf 657}, 55 (2007)
  [arXiv:0707.2452 [hep-ph]].
  %%CITATION = ARXIV:0707.2452;%%
  %10 citations counted in INSPIRE as of 16 Sep 2014


%\cite{Becher:2006nr}
\bibitem{Becher:2006nr}
  T.~Becher and M.~Neubert,
  %``Threshold resummation in momentum space from effective field theory,''
  Phys.\ Rev.\ Lett.\  {\bf 97}, 082001 (2006)
  [hep-ph/0605050].
  %%CITATION = HEP-PH/0605050;%%
  %87 citations counted in INSPIRE as of 16 Sep 2014


%\cite{Becher:2007ty}
\bibitem{Becher:2007ty}
  T.~Becher, M.~Neubert and G.~Xu,
  %``Dynamical Threshold Enhancement and Resummation in Drell-Yan Production,''
  JHEP {\bf 0807}, 030 (2008)
  [arXiv:0710.0680 [hep-ph]].
  %%CITATION = ARXIV:0710.0680;%%
  %125 citations counted in INSPIRE as of 16 Sep 2014


%\cite{Ahrens:2008nc}
\bibitem{Ahrens:2008nc}
  V.~Ahrens, T.~Becher, M.~Neubert and L.~L.~Yang,
  %``Renormalization-Group Improved Prediction for Higgs Production at Hadron Colliders,''
  Eur.\ Phys.\ J.\ C {\bf 62}, 333 (2009)
  [arXiv:0809.4283 [hep-ph]].
  %%CITATION = ARXIV:0809.4283;%%
  %119 citations counted in INSPIRE as of 16 Sep 2014


%\cite{Bonvini:2012az}
\bibitem{Bonvini:2012az}
  M.~Bonvini, S.~Forte, M.~Ghezzi and G.~Ridolfi,
  %``Threshold Resummation in SCET vs. Perturbative QCD: An Analytic Comparison,''
  Nucl.\ Phys.\ B {\bf 861}, 337 (2012)
  [arXiv:1201.6364 [hep-ph]].
  %%CITATION = ARXIV:1201.6364;%%
  %10 citations counted in INSPIRE as of 16 Sep 2014


%\cite{Sterman:2013nya}
\bibitem{Sterman:2013nya}
  G.~Sterman and M.~Zeng,
  %``Quantifying Comparisons of Threshold Resummations,''
  JHEP {\bf 1405}, 132 (2014)
  [arXiv:1312.5397 [hep-ph]].
  %%CITATION = ARXIV:1312.5397;%%
  %4 citations counted in INSPIRE as of 16 Sep 2014


%\cite{Almeida:2014uva}
\bibitem{Almeida:2014uva}
  L.~G.~Almeida, S.~D.~Ellis, C.~Lee, G.~Sterman, I.~Sung and J.~R.~Walsh,
  %``Comparing and counting logs in direct and effective methods of QCD resummation,''
  JHEP {\bf 1404}, 174 (2014)
  [arXiv:1401.4460 [hep-ph]].
  %%CITATION = ARXIV:1401.4460;%%
  %5 citations counted in INSPIRE as of 16 Sep 2014


%\cite{Bonvini:2014qga}
\bibitem{Bonvini:2014qga}
  M.~Bonvini, S.~Forte, G.~Ridolfi and L.~Rottoli,
  %``Resummation prescriptions and ambiguities in SCET vs. direct QCD: Higgs production as a case study,''
  arXiv:1409.0864 [hep-ph].
  %%CITATION = ARXIV:1409.0864;%%


%\cite{Becher:2006mr}
\bibitem{Becher:2006mr}
  T.~Becher, M.~Neubert and B.~D.~Pecjak,
  %``Factorization and Momentum-Space Resummation in Deep-Inelastic Scattering,''
  JHEP {\bf 0701}, 076 (2007)
  [hep-ph/0607228].
  %%CITATION = HEP-PH/0607228;%%
  %124 citations counted in INSPIRE as of 16 Sep 2014


%\cite{Lai:2010vv}
\bibitem{Lai:2010vv}
  H.~L.~Lai, M.~Guzzi, J.~Huston, Z.~Li, P.~M.~Nadolsky, J.~Pumplin and C.-P.~Yuan,
  %``New parton distributions for collider physics,''
  Phys.\ Rev.\ D {\bf 82}, 074024 (2010)
  [arXiv:1007.2241 [hep-ph]].
  %%CITATION = ARXIV:1007.2241;%%
  %992 citations counted in INSPIRE as of 16 Sep 2014


%\cite{Gao:2013xoa}
\bibitem{Gao:2013xoa}
  J.~Gao, M.~Guzzi, J.~Huston, H.~L.~Lai, Z.~Li, P.~Nadolsky, J.~Pumplin and D.~Stump {\it et al.},
  %``The CT10 NNLO Global Analysis of QCD,''
  Phys.\ Rev.\ D {\bf 89}, 033009 (2014)
  [arXiv:1302.6246 [hep-ph]].
  %%CITATION = ARXIV:1302.6246;%%
  %111 citations counted in INSPIRE as of 16 Sep 2014


%\cite{Buchalla:1995vs}
\bibitem{Buchalla:1995vs}
  G.~Buchalla, A.~J.~Buras and M.~E.~Lautenbacher,
  %``Weak decays beyond leading logarithms,''
  Rev.\ Mod.\ Phys.\  {\bf 68}, 1125 (1996)
  [hep-ph/9512380].
  %%CITATION = HEP-PH/9512380;%%
  %1821 citations counted in INSPIRE as of 16 Sep 2014

\end{thebibliography}
\end{document}